\documentclass[journal]{IEEEtran}
\usepackage{slashbox, graphicx, times, amsmath, amssymb, cite, subfigure, stfloats, bm}
\usepackage[lined,ruled,commentsnumbered]{algorithm2e}
\usepackage{booktabs,multirow}

\makeatletter

\newcommand{\Rmnum}[1]{\expandafter\@slowromancap\romannumeral #1@}
\makeatother

\begin{document}
\title{Multi-Task Deep Learning with Dynamic Programming for Embryo Early Development Stage Classification from Time-Lapse Videos}
\author{Zihan~Liu, Bo~Huang, Yuqi~Cui, Yifan~Xu, Bo~Zhang, Lixia~Zhu, Yang~Wang, Lei~Jin and Dongrui~Wu
\thanks{Z.~Liu, Y.~Cui, Y.~Xu, Y.~Wang and D.~Wu are with the Key Laboratory of the Ministry of Education for Image Processing and Intelligent Control, School of Artificial Intelligence and Automation, Huazhong University of Science and Technology, Wuhan 430074, China. Email: zhliu95@hust.edu.cn, yqcui@hust.edu.cn, yfxu@hust.edu.cn, wangyang\_sky@hust.edu.cn, drwu@hust.edu.cn.}
\thanks{B.~Huang, B.~Zhang, L.~Zhu and L.~Jin are with the Reproductive Medicine Center, Tongji Hospital, Tongji Medical College, Huazhong University of Science and Technology, Wuhan 430074, China. Email: hb2627@gmail.com, dramanda@126.com, zhulixia027@foxmail.com, jinleirepro@yahoo.com.}
\thanks{Zihan~Liu and Bo~Huang contributed equally to this work.}
\thanks{Lei~Jin and Dongrui~Wu are the corresponding authors. }}

\maketitle

\begin{abstract}
Time-lapse is a technology used to record the development of embryos during in-vitro fertilization (IVF). Accurate classification of embryo early development stages can provide embryologists valuable information for assessing the embryo quality, and hence is critical to the success of IVF. This paper proposes a multi-task deep learning with dynamic programming (MTDL-DP) approach for this purpose. It first uses MTDL to pre-classify each frame in the time-lapse video to an embryo development stage, and then DP to optimize the stage sequence so that the stage number is monotonically non-decreasing, which usually holds in practice. Different MTDL frameworks, e.g., one-to-many, many-to-one, and many-to-many, are investigated. It is shown that the one-to-many MTDL framework achieved the best compromise between performance and computational cost. To our knowledge, this is the first study that applies MTDL to embryo early development stage classification from time-lapse videos.
\end{abstract}


\IEEEpeerreviewmaketitle

\section{Introduction} \label{sect:Intro}

In-vitro fertilization (IVF) \cite{Huang2016, Huang2014a, Huang2015a} is a frequently used technology for treating infertility. The process involves the collection of multiple follicles for fertilization and in-vitro culture. Cultivation, selection and transplantation of embryo are the key steps in determining a successful implantation during IVF \cite{ReproductiveMedicine2011,Tomasz2004}. During the development of embryos, the morphological characteristics \cite{holte2006construction} and kinetic characteristics \cite{lemmen2008kinetic} are highly correlated with the outcome of transplantation.

Time-lapse videos have been widely used in various reproductive medicine centers during the cultivation of embryos \cite{Kirkegaard2012} to monitor them. A time-lapse video records the embryonic development process in real time by taking photos of the embryos at short time intervals \cite{Wong2010}. Thus, a large amount of time series image data for each embryo are produced in this process. At the final stage of embryo selection, an embryologist reviews the entire embryo development process to score and sort them. Studies with different time-lapse equipment reported improved prediction accuracy of embryo implantation potential by analyzing the morphokinetics of human embryos at early cleavage stages \cite{Wong2010,Herrero2013,Chen2013,Kirkegaard2012,Meseguer2011}. These features have been shown to be statistically significant to the final outcome of the transplantation \cite{lemmen2008kinetic}.

There have been only a few approaches to analyze time-lapse image data \cite{Wong2010,Wang2013c,Conaghan2013,jonaitis2016application,khan2016segmentation,Ng_2018_ICLR,khan2016deep}. Due to the limitations of the time-lapse technology, stereoscopic cells of different heights overlap in the images when photographed. It is difficult for even an experienced embryologist to accurately count the number of cells in a single time-lapse image when there are more than eight cells. Therefore, most research focused on the early development stages of embryos. Wong \emph{et al.} \cite{Wong2010} identified several key parameters that can predict blastocyst formation at the 4-cell stage from time-lapse images, and employed sequential Monte Carlo based probabilistic model estimation to monitor these parameters and track the cells. Wang \emph{et al.} \cite{Wang2013c} presented a multi-level embryo stage classification approach, by using both hand-crafted and automatically learned embryo features to identify the number of cells in a time-lapse video. Conaghan \emph{et al.} \cite{Conaghan2013} used an automated and proprietary image analysis software EEVA\textsuperscript{TM} (Early Embryo Viability Assessment), which exhibited high image contrast through the use of darkfield illumination, to track cell divisions from one-cell stage to four-cell stage. Their experiments verified that the EEVA Test can significantly improve embryologists' ability to identify embryos that would develop into usable blastocysts. There are also several other studies on embryo selection by using EEVA\textsuperscript{TM} \cite{VerMilyea2014,Diamond2015,Aparicio-Ruiz2016,Kieslinger2016}, but they did not provide the details of the used EEVA Test. Jonaitis \emph{et al.} \cite{jonaitis2016application} compared the performance of neural network, support vector machine and nearest neighbor classifier in detecting cell division time. Khan \emph{et al.} \cite{khan2016deep} used a deep convolutional neural network (CNN) to classify the number of cells, and also semantic segmentation to extract the cell regions in a time-lapse image \cite{khan2016segmentation}. Ng \emph{et al.} \cite{Ng_2018_ICLR} combined late fusion networks with dynamic programming (DP) to predict different cell development stages and obtained better results than a single-frame model.

Multi-task learning has been successfully used in many applications, such as natural language processing \cite{Collobert2008}, speech recognition \cite{Deng2013}, and computer vision \cite{Girshick_2015}. Its basic idea is to share representations among related tasks, so that each trained model may have better generalization ability \cite{Ruder17a}. This paper proposes a multi-task deep learning with dynamic programming (MTDL-DP) approach, which first uses MTDL to pre-classify each frame in the time-lapse video to an embryo development stage, and then DP to optimize the stage sequence so that the stage number is monotonically non-decreasing, which usually holds in practice. To our knowledge, this is the first study that applies MTDL to embryo early development stage classification from time-lapse videos.

The remainder of this paper is organized as follows: Section~\ref{sect:Method} introduces four classification frameworks for time-lapse video analysis. Section~\ref{sect:MTDL-DP} proposes our MTDL-DP approach. Section~\ref{sect:experiments} presents the experimental results. Finally, Section~\ref{sect:conclusions} draws conclusion.

\section{Classification Frameworks} \label{sect:Method}

This section introduces four frameworks for embryo early development stage classification from time-lapse videos. We first describe our dataset and the baseline network architecture, and then extend it to \emph{many-to-one}, \emph{one-to-many} and \emph{many-to-many} MTDL frameworks.

\subsection{Dataset}

\begin{figure*}[htpb] \centering
\subfigure[]{\includegraphics[width=0.45\linewidth,clip]{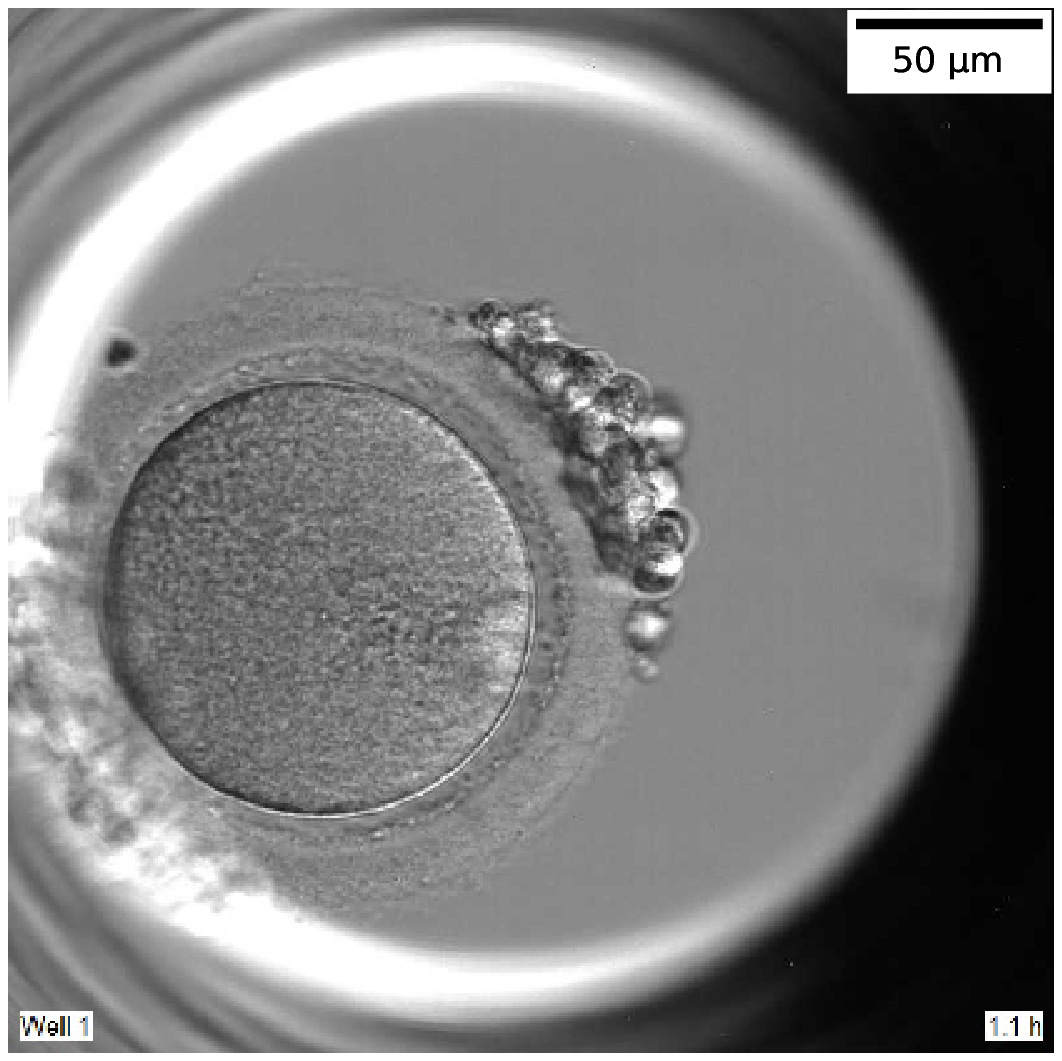}}
\subfigure[]{\includegraphics[width=0.45\linewidth,clip]{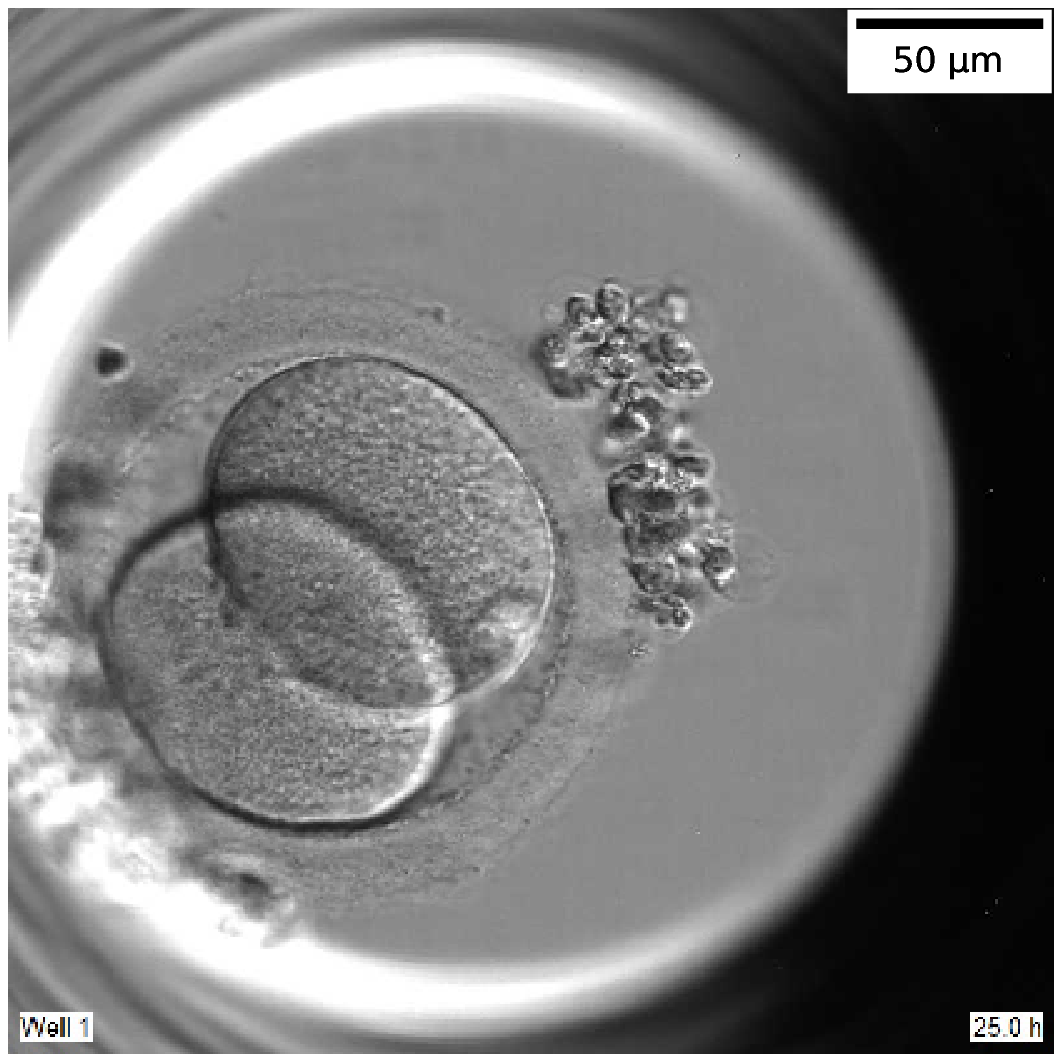}}
\subfigure[]{\includegraphics[width=0.45\linewidth,clip]{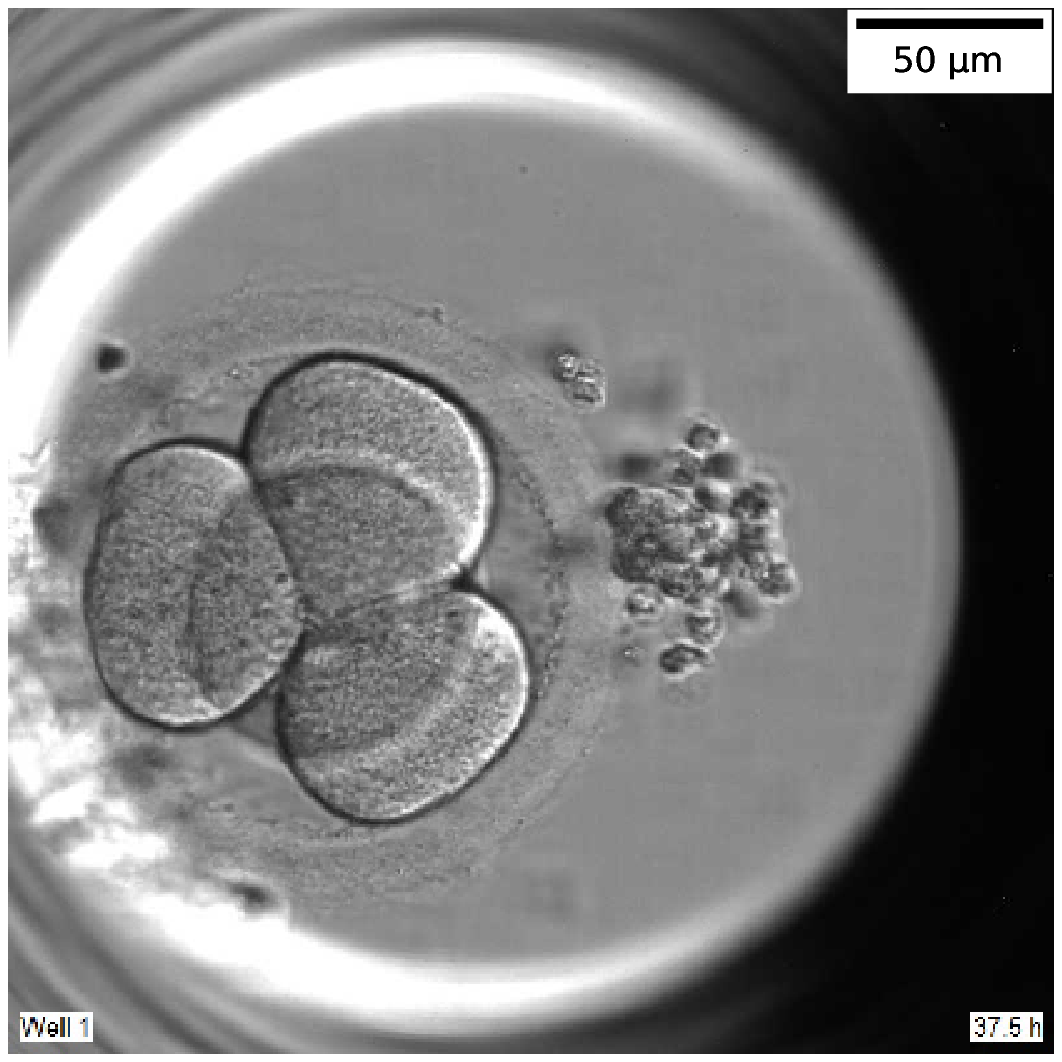}}
\subfigure[]{\includegraphics[width=0.45\linewidth,clip]{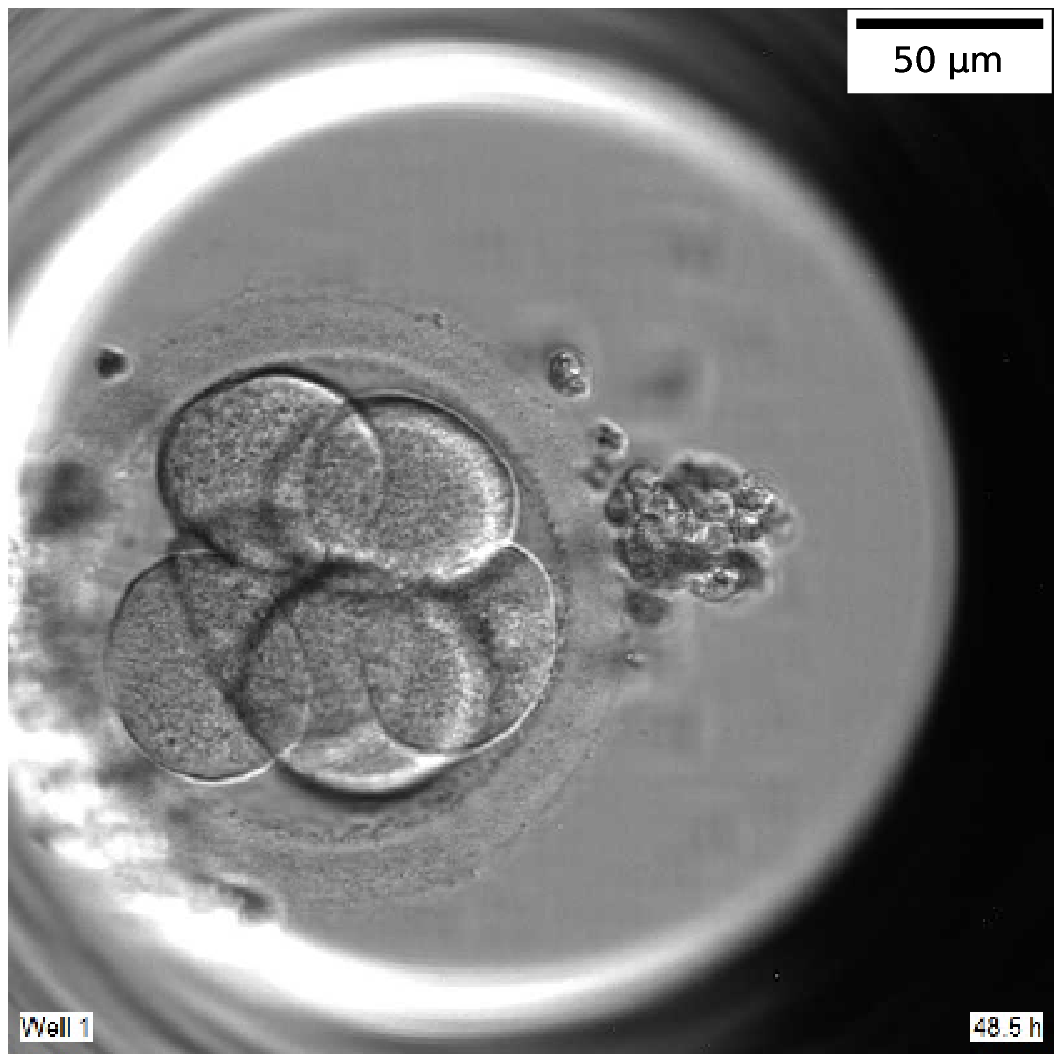}}
\caption{Sample frames from a time-lapse video. (a)~1-cell stage; (b)~2-cell stage; (c)~4-cell stage; (d)~4+-cell stage.} \label{fig:cell_image}
\end{figure*}

The time-lapse video dataset used in our experiments came from the Reproductive Medicine Center of Tongji Hospital, Huazhong University of Science and Technology, Wuhan, China. It consisted of 170 time-lapse videos extracted from incubators, using an EmbryoScope+ time-lapse microscope system\footnote{https://www.vitrolife.com/products/time-lapse-systems/embryoscopeplus-time-lapse-system/} at 10-minute sampling interval. Each frame in the video is a grayscale $800\times800$ image with a well number in the lower left corner and a time marker after fertilization in the lower right corner, as shown in Fig.~\ref{fig:cell_image}. The embryo is surrounded by some granulosa cells in the microscope field. The scale bar in the upper right corner indicates the size of the cells. Each video began about 0-2 hours after fertilization, and ended about 140 hours after fertilization. We only used the first $N=350$ frames in each video, which were manually labeled for the embryo development stages. Therefore, we had a total of $170\times350=59,500$ labeled frames in the experiment.

\begin{figure}[htpb]\centering
\includegraphics[width=\linewidth,clip]{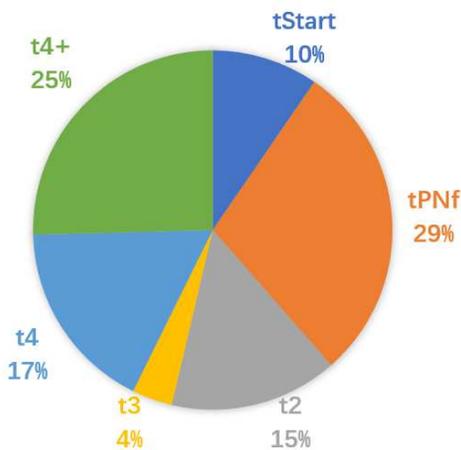}
\caption{Percentage of frames in different embryo development stages.} \label{fig:distribution}
\end{figure}

As in \cite{Ng_2018_ICLR}, we focused on the first six embryo development stages, which included initialization (tStart), the appearance and breakdown of the male and female pronucleus (tPNf), and the appearance of 2 through 4+ cells (t2, t3, t4, t4+). We counted the number of images in different embryo development stages in the dataset, and show the summary in Fig.~\ref{fig:distribution}. Note that t3 was rarely observed in our dataset.

\subsection{The Baseline One-to-One Classification Framework}

Let $\textbf{x}_n$ be the $n$th frame in a time-lapse video. For image classification, a standard \emph{one-to-one} classification framework learns a mapping:
\begin{align}
f_0: \textbf{x}_{n} \mapsto y_{n} \in \emph{L}
\end{align}
where $y_n$ is the stage label of $\textbf{x}_n$, and $L$ the label set of the embryo development stages.

\begin{figure*}[htpb]\centering
\subfigure[]{ \includegraphics[width=0.15\linewidth,clip]{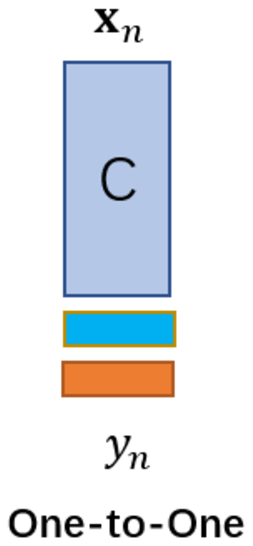}} \qquad \qquad
\subfigure[]{\label{fig:manyOne} \includegraphics[width=0.55\linewidth,clip]{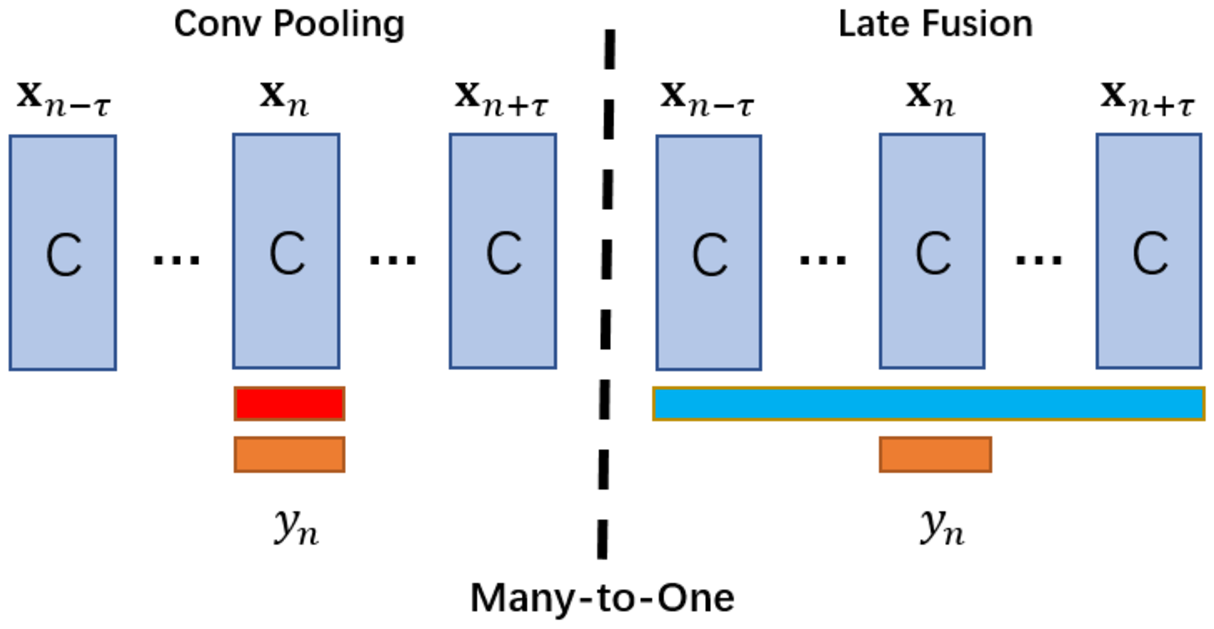}} \\
\subfigure[]{\label{fig:oneMany}  \includegraphics[width=0.28\linewidth,clip]{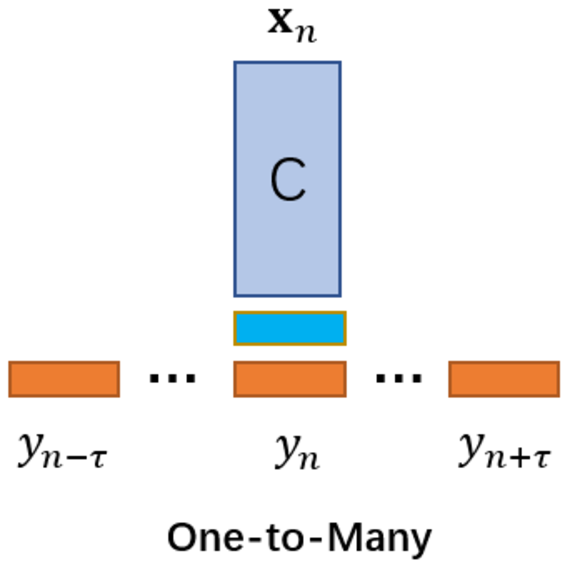}} \qquad \qquad
\subfigure[]{\label{fig:manyMany} \includegraphics[width=0.28\linewidth,clip]{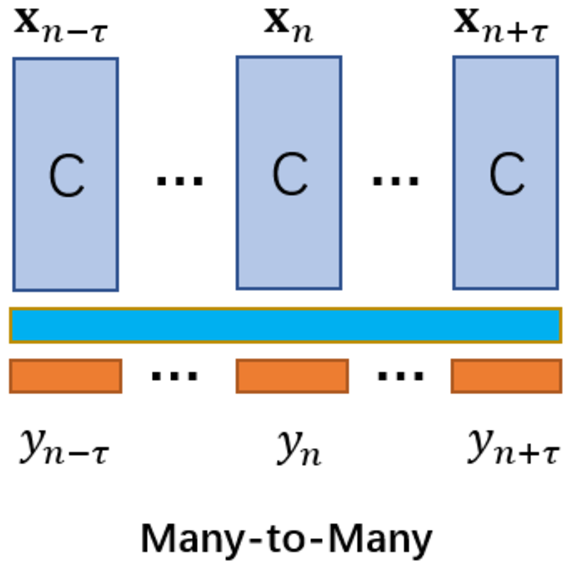}}\\
\caption{Different classification frameworks. (a)~\emph{one-to-one}; (b)~\emph{many-to-one}; (c)~\emph{one-to-many}; (d)~\emph{many-to-many}. The convolutional layers are denoted by `C'. Blue and red rectangles denote the flatten layer and the max-pooling layer, respectively. Orange rectangles denote the fully connected and softmax layers.} \label{fig:frameworks}
\end{figure*}

When information of the previous and future frames is used, the standard \emph{one-to-one} classification framework can be extended to \emph{many-to-one}, \emph{one-to-many} and \emph{many-to-many} MTDL frameworks, as illustrated in Fig.~\ref{fig:frameworks}.

We used ResNet \cite{He2016}, which won the 2015 ImageNet classification competition, to process individual video frames. Table~\ref{tab:ResNet50} shows our baseline ResNet50 model. The input image had three channels (RGB), each with 224$\times$224 pixels (the 800$\times$800 images were resized). The model was initialized from the ResNet weights pre-trained on ImageNet \cite{Deng2009}, which can help reduce overfitting on small datasets.

\begin{table}[htbp] \center
\caption{The baseline ResNet50 model.} \label{tab:ResNet50}
\linespread{1.35}\selectfont
\begin{tabular}{c|c|c}
\hline
layer name & 50-layer & output size \\ \hline
conv1                 & $7\times7$, 64, stride $2$       & $112\times112\times64$                 \\ \hline
\multirow{2}{*}{res2} & $3\times3$ max pool, stride $2$ & \multirow{4}{*}{$56\times56\times256$} \\ \cline{2-2}
& $
\left[ \begin{array}{lc}
1 \times 1,\;64\\
3 \times 3,\;64\\
1 \times 1,\;256
\end{array} \right] \times 3\
$ & \\ \hline
res3 & $
\left[ \begin{array}{lc}
1 \times 1,\;128\\
3 \times 3,\;128\\
1 \times 1,\;512
\end{array} \right] \times 4\
$ & $28\times28\times512$ \\ \hline
res4 & $
\left[ \begin{array}{lc}
1 \times 1,\;256\\
3 \times 3,\;256\\
1 \times 1,\;1024
\end{array} \right] \times 6\
$ & $14\times14\times1024$ \\ \hline
res5 & $
\left[ \begin{array}{lc}
1 \times 1,\;512\\
3 \times 3,\;512\\
1 \times 1,\;2048
\end{array} \right] \times 3\
$ & $7\times7\times2048$ \\ \hline
\multicolumn{2}{c|}{global average pool} & $1\times1\times2048$ \\ \hline
\multicolumn{2}{c|}{fc}                       & $1\times1\times1024$                   \\ \hline
\multicolumn{2}{c|}{softmax}                  & $1\times1\times6$                      \\ \hline
\end{tabular}
\end{table}

\subsection{The Many-to-One MTDL Framework}

The \emph{many-to-one} MTDL framework, shown in Fig.~\ref{fig:manyOne}, is frequently used in video understanding \cite{Soomro2012,Karpathy_2014_CVPR,CabaHeilbron2015} because multiple frames in the same video usually have the same label, and hence they can be considered together to predict the final label. \emph{Many-to-one} can better make use of input context information than \emph{one-to-one}.

\emph{Many-to-one} performs the following mapping:
\begin{align}
f_1: (\textbf{x}_{n-\tau},\ldots,\textbf{x}_{n},\ldots,\textbf{x}_{n+\tau}) \mapsto y_{n} \in \emph{L},
\end{align}
where $\tau$ is the number of neighboring frames before and after the current frame (the input context window size is hence $2\tau+1$).

There are two common approaches to fuse time domain information from the $2\tau+1$ frames: Conv Pooling \cite{Ng2015} and Late Fusion \cite{Karpathy_2014_CVPR}.

\subsubsection{Conv Pooling}

This is a convolutional temporal feature pooling architecture, which has been extensively used for video classification, especially for bag-of-words representations \cite{Fei2005}. Image features are computed for each frame and then max pooled. The pooling features can then be sent to fully connected layers for final classification. A major advantage of this approach is that spatial information in multiple frames, output by the convolutional layers, is preserved through a max pooling operation in the time domain. Experiments \cite{Ng2015} verified that Conv Pooling outperformed all other feature pooling approaches on the Sports-1M dataset, using a 120-frame AlexNet model \cite{Krizhevsky2012}.

\subsubsection{Late Fusion}

In Late Fusion, all frames in the input context window are encoded via identical ConvNets. The final representations after all convolutional layers are concatenated and passed through a fully-connected layer to generate classifications. The concatenation can happen to a subset of frames in the input context window \cite{Karpathy_2014_CVPR}, or to all frames in that window \cite{Ng_2018_ICLR}. Previous research \cite{Ng_2018_ICLR} demonstrated that Late Fusion ConvNets using 15 frames and a DP-based decoder outperformed Early Fusion for predicting embryo morphokinetics in time-lapse videos.

\subsection{The One-to-Many MTDL Framework} \label{sect:o2m}

\begin{figure}[htpb] \centering
\includegraphics[width=0.95\linewidth,clip]{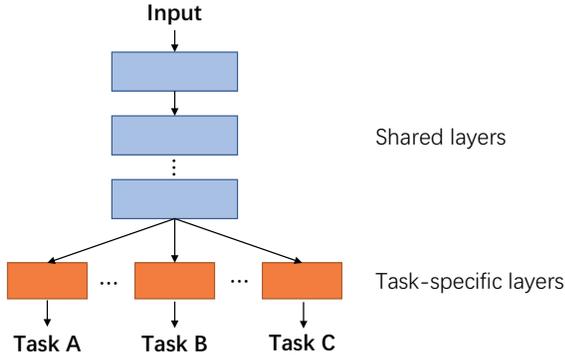}
\caption{Hard parameter sharing for MTDL.} \label{fig:hard_sharing}
\end{figure}

\emph{One-to-many}, shown in Fig.~\ref{fig:oneMany}, means each input is mapped to multiple outputs, which is also called multi-task nets \cite{Dahl2014} in deep learning. This paper uses \emph{hard parameter sharing} of hidden layers \cite{Ruder17a}, as illustrated in Fig.~\ref{fig:hard_sharing}. The parameters of the convolution layers are shared among different tasks, but those of the fully connected layers are trained separately.

In \emph{one-to-many}, each $\mathbf{x}_n$ is used in classifying $2\tau+1$ stages centered at $n$, i.e., it learns the following one-to-many mapping:
\begin{align}
f_2: \textbf{x}_n \mapsto (y_{n-\tau},\ldots,y_{n+\tau}) \in \emph{L}^{2\tau+1}.
\end{align}
$\mathbf{x}_n$'s classification for the stage at time index $t\in[n-\tau,n+\tau]$ is a probability vector $\hat{\bm{p}}_t(\bm{x}_n)\in \mathbb{R}^{|L|\times1}$.

At each Frame Index $n$, the corresponding label is estimated by $2\tau+1$ neighboring $\mathbf{x}_t$, $t\in[n-\tau,n+\tau]$. We need to aggregate them to obtain the final classification. This can be done by an ensemble approach.

Because each frame $\mathbf{x}_n$ is involved in $2\tau+1$ outputs, the total loss on a training frame $\mathbf{x}_n$ is computed as the sum of the loss on all involved outputs:
\begin{align}
\ell(\mathbf{x}_n)=\sum^{n+\tau}_{t=n-\tau}w_t\cdot\ell(y_t,\hat{\bm{p}}_{t}(\mathbf{x}_n)),
\end{align}
where $w_t$ is the weight for the $t$-th output, and $y_t$ is the true label for Frame $t$. $w_t=1$ and the cross-entropy loss were used in this paper. The cross-entropy loss on the $t$-th output can be written as follows:
\begin{align}
\ell(y_t,\hat{\bm{p}}_{t}(\mathbf{x}_n))=-\log{(\hat{p}_{t,y_t}(\mathbf{x}_n))},
\end{align}
where $\hat{p}_{t,y_t}(\mathbf{x}_n)$ is the $y_t$-th element of $\hat{\bm{p}}_t(\mathbf{x}_n)$.

\subsection{The Many-to-Many MTDL Framework}

\emph{Many-to-many} can be viewed as a combination of \emph{one-to-many} and \emph{many-to-one}. Each input frame is processed by a separate CNN. Late Fusion was used, and the parameters of the fully connected layers were also trained separately, as shown in Fig.~\ref{fig:manyMany}.

\section{Multi-Task Deep Learning with Dynamic Programming (MTDL-DP)} \label{sect:MTDL-DP}

This section introduces our proposed MTDL-DP approach.

\subsection{Ensemble Learning for MTDL} \label{sect:ensemble}

As mentioned in Section~\ref{sect:o2m}, a multi-task net has multiple outputs. The easiest approach to get the final classification corresponding to a specific frame is to choose the middle output of the network. A more sophisticated approach is ensemble learning \cite{Zhou2012}. We consider two common probabilistic aggregation approaches in this paper: \emph{additive mean} and \emph{multiplicative mean}.

\begin{figure*}[htpb] \centering
\includegraphics[width=0.9\linewidth,clip]{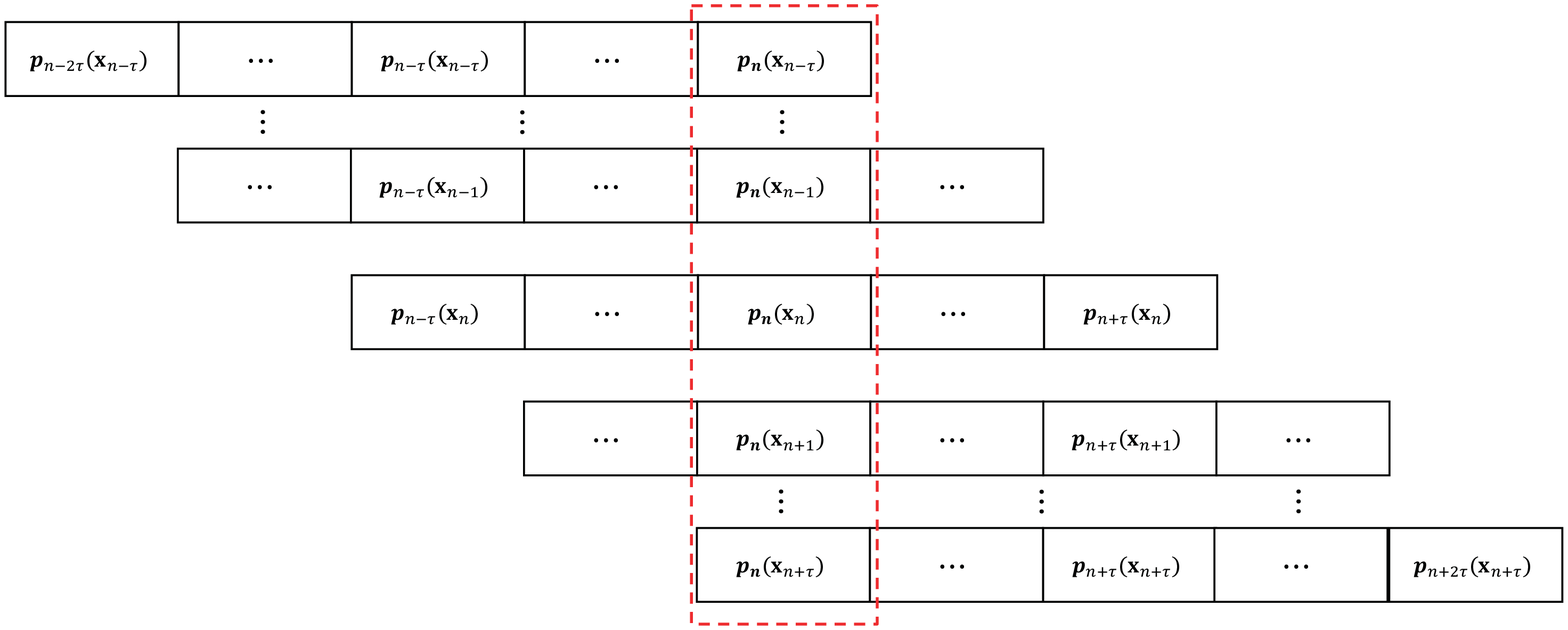}
\caption{Ensemble of the multi-task net's predictions at Frame Index $n$, made by neighboring frames $\textbf{x}_t$, $t\in [n-\tau,n+\tau]$.} \label{fig:ensemble_frame}
\end{figure*}

Let $\hat{\bm{p}}_{n}(\textbf{x}_t)$ be the predicted probability vector at Frame Index $n$, given Frame $\textbf{x}_t$, $t\in[n-\tau,n+\tau]$, as illustrated in Fig.~\ref{fig:ensemble_frame}. The ensemble probability $\hat{\bm{p}}_n$ at Frame Index $n$, aggregated by the additive mean, is:
\begin{align}
\hat{\bm{p}}_{n}=\frac{1}{2\tau+1} \sum^{n+\tau}_{t=n-\tau}\hat{\bm{p}}_n(\textbf{x}_t), \label{eq:p_a}
\end{align}
If the multiplicative mean is used,
\begin{align}
\hat{\bm{p}}_{n}=\frac{1}{2\tau+1} \prod^{n+\tau}_{t=n-\tau}\hat{\bm{p}}_n(\textbf{x}_t). \label{eq:p_b}
\end{align}
Since each $\hat{\bm{p}}_n(\textbf{x}_t)$ is a vector, the summation in (\ref{eq:p_a}) and multiplication in (\ref{eq:p_b}) are element-wise operations.

The final classification label $\hat{y}_n$ for Frame $\mathbf{x}_n$ is obtained by probability maximization:
\begin{align}
\hat{y}_n=\mathop{\arg\max}_{1\le l\le |L|}\hat{p}_{n,l},
\end{align}
where $\hat{p}_{n,l}$ is the $l$-th element of $\hat{\bm{p}}_n$.

\subsection{Post-processing with DP} \label{sect:dp}

The number of cells in the development of an embryo is almost always non-decreasing \cite{Liu2014}. However, this is not guaranteed in the classification outputs of MTDL. We use DP to adjust the classifications so that this constraint is satisfied.

For each video, the groundtruth stages $\{y_n\}_{n=1}^N$ form a sequence. MTDL outputs a probability vector $\hat{\bm{p}}_n=[p_{n,1},...,p_{n,|L|}]^T$ before likelihood maximization at Frame $n$, where $\hat{p}_{n,l}$ is the estimated probability that Frame $n$ is at Stage $l$. We define $E(\bm{\hat{y}},\mathbf{\hat{P}})$ as the total loss for an estimated prediction $\bm{\hat{y}}=\{\hat{y}_n\}_{n=1}^{N}$, given the model output probability matrix $\mathbf{\hat{P}}=[\hat{\bm{p}}_1,...,\hat{\bm{p}}_N]$. The total loss is the sum of the per-frame losses $\sum_{n=1}^{N}{e(\hat{y}_n,\hat{\bm{p}}_n)}$, which must be optimized subject to the monotonicity constraint: $\hat{y}_{n+1}\geq{\hat{y}_{n}},\ \forall{n}$.

Two common per-frame losses \cite{Ng_2018_ICLR} were used. The first is negative label likelihood (LL), defined as:
\begin{align} \label{eq:e_ll}
e_{LL}(\hat{y}_n,\hat{\bm{p}}_n)=-\log{(\hat{p}_{n,y_n})}.
\end{align}
The second is earth mover (EM) distance, defined as:
\begin{align} \label{eq:e_em}
e_{EM}(\hat{y}_n,\hat{\bm{p}}_n)=-\sum_{l=1}^{|L|}{\hat{p}_{n,l}\vert \hat{y}_n-l \vert},
\end{align}

The final classification stage sequence $\bm{\hat{y}}^*=\{\hat{y}_n\}_{n=1}^{N}$ can be obtained as:
\begin{align} \label{eq:video_y}
&\hat{\bm{y}}^*=\mathop{\arg\min}_{\bm{\hat{y}}=\{\hat{y}_n\}_{n=1}^{N}}\sum_{n=1}^{N}{e(\hat{y}_n,\hat{\bm{p}}_n)} \\ \nonumber
&\text{s.t.} \quad \hat{y}_{n+1}\geq{\hat{y}_{n}},\ \forall{n}.
\end{align}
which can be easily solved by DP, as shown in Algorithm~\ref{alg:DP}.

\begin{algorithm}[h] %\DontPrintSemicolon
\KwIn{$N$, the number of frames in a time-lapse video\;
$L$, label set of embryo development stages\;
$\mathbf{\hat{P}}=[\hat{\bm{p}}_1,...,\hat{\bm{p}}_N]\in\mathbb{R}^{|L|\times N}$, the MTDL model output probability matrix for the $N$ frames.}
\KwOut{$\bm{\hat{y}}^*$, the optimized stage sequence.}

Set $e(l,\hat{\bm{p}}_n)=0$ and $E(l,\hat{\bm{p}}_n)=0$, $\forall l\in[1,|L|]$, $\forall n\in[1,N]$\;
\For{$n=1,...,N$}{
\For{$\hat{y}=1,...,|L|$}{
Compute $e(\hat{y},\hat{\bm{p}}_n)$ in (\ref{eq:e_em})\;}
}
\For{$n=2,...,N$}{
\For{$\hat{y}=1,...,|L|$}{
$\displaystyle E(\hat{y},\hat{\bm{p}}_n)=e(\hat{y},\hat{\bm{p}}_n)+\min_{1\le l \le\hat{y}} E(l,\hat{\bm{p}}_{n-1})$\;
}
}
$k=|L|$\;
\For{$n=N,...,1$}{
$\displaystyle \hat{y}_n=\mathop{\arg\min}_{1\leqslant{l}\leqslant{k}}{E(l,\hat{\bm{p}}_n)}$\;
\If{$\hat{y}_n<k$}{
$k=\hat{y}_n$\;
}
}
$\bm{\hat{y}}^*=\{\hat{y}_n\}_{n=1}^{N}$\;
\textbf{Return} The optimized stage sequence $\bm{\hat{y}}^*$.
\caption{Pseudocode of dynamic programming (DP).} \label{alg:DP}
\end{algorithm}

\subsection{MTDL-DP}

Our proposed MTDL-DP consists of three steps: 1) construct a multi-task net with the \emph{one-to-many} or \emph{many-to-many} MTDL framework; 2) use multiplicative mean to aggregate the prediction of the multi-task net; and, 3) post-process with DP using the EM distance per-frame loss. Its pseudocode is given in Algorithm~\ref{alg:MTDL-DP}.

\begin{algorithm}[h] %\DontPrintSemicolon
\KwIn{$N$, the number of frames in a time-lapse video\;
$D$, set of labeled time-lapse videos\;
$\{\mathbf{x}_n\}_{n=1}^N$, frames to be labeled\;
$\tau$, the number of left and right neighboring frames in the context window.}
\KwOut{$\bm{\hat{y}}^*$, the labeled stage sequence.}

Use the \emph{one-to-one} framework to train a baseline model $f_0$ from $D$\;
Initialize an MTDL model, whose convolution layer parameters are identical to $f_0$\;
Fine-tune the fully connected layer parameters of the MTDL model on $D$\;
\For{$n=1,...,N$}{
Use the MTDL model to compute $\hat{\bm{p}}_t(\mathbf{x}_n)$, $t=n-\tau,...,n+\tau$\;
}
\For{$n=1,...,N$}{
Compute $\hat{\bm{p}}_n$ by (\ref{eq:p_b})\;
Compute the per-frame loss $e_{EM}(\hat{y}_n,\hat{\bm{p}}_n)$ in (\ref{eq:e_em})\;
}
Solve for $\bm{\hat{y}}^*$ in (\ref{eq:video_y}) by Algorithm~1\;

\textbf{Return} The optimized stage sequence $\bm{\hat{y}}^*$.
\caption{MTDL-DP} \label{alg:MTDL-DP}
\end{algorithm}

The \emph{one-to-many} MTDL framework can also be replaced by the \emph{many-to-many} MTDL framework.

\section{Experimental Results} \label{sect:experiments}

This section investigates the performance of our proposed MTDL-DP.

\subsection{Experimental Setup}

We created training/validation/test data partitions by randomly selecting 70\%/10\%/20\% videos from the dataset, i.e., 41,650/5,950/11,900 frames, respectively. We resized each frame to 224$\times$224 so that it can be used by ResNet50, our baseline model. Random rotation and flip data augmentation was used. All MTDL frameworks were initialized by the weights trained by \emph{one-to-one} (ResNet50). Then, the convolution layer parameters were frozen, and the fully connected layers were further tuned.

We used the cross-entropy loss function and Adam optimizer \cite{Adam}, and early stopping to reduce overfitting, in all experiments. Multiplicative mean and EM distance per-frame loss were used in the MTDL-DP. All experiments were repeated five times, and the mean results were reported.

\subsection{Classification Accuracy}

First, we considered MTDL only, without using DP. The classification accuracies are shown in the left panel of Table~\ref{tab:acc1}, with $\tau=\{1, 4, 7\}$ (the output context window size was $2\tau+1$). All MTDL frameworks outperformed the \emph{one-to-one} framework, suggesting using neighboring input or label information in multi-task learning was indeed beneficial.

\begin{table*}[htbp] \center
\linespread{1.35}\selectfont
\caption{Classification accuracies for different classification frameworks and $\tau$, before and after DP post-processing.}   \label{tab:acc1}
\begin{tabular}{|c|c|c|c|c|c|c|c|}
\hline
\multirow{2}{*}{Framework}   & \multirow{2}{*}{Method}          & \multicolumn{3}{c|}{Accuracy without DP}                & \multicolumn{3}{c|}{Accuracy with DP}                 \\ \cline{3-8}
                             &                                  & $\tau=1$        & $\tau=4$        & $\tau=7$        & $\tau=1$        & $\tau=4$        & $\tau=7$        \\ \hline
One-to-one                   & ResNet50                         & 83.8\%          & 83.8\%          & 83.8\%          & 86.1\%          & 86.1\%          & 86.1\%          \\ \hline
\multirow{2}{*}{Many-to-one} & Conv Pooling                     & 84.7\%          & 84.4\%          & 83.8\%          & 85.9\%          & 85.1\%          & 84.5\%          \\ \cline{2-8}
                             & Late Fusion                      & 83.9\%          & 84.6\%          & 85.1\%          & 86.0\%          & 85.2\%          & 85.2\%          \\ \hline
One-to-many                  & \multirow{2}{*}{Multi-Task Nets (ours)} & \textbf{85.0\%}  & 85.4\%  & 85.3\%          & 86.5\%          & 85.8\%          & 85.7\%          \\ \cline{1-1} \cline{3-8}
Many-to-many                 &                                  & 84.6\%          & \textbf{85.7\%} & \textbf{85.8\%} & \textbf{86.6\%} & \textbf{86.5\%} & \textbf{86.9\%} \\ \hline
\end{tabular}
\end{table*}

For the \emph{many-to-one} MTDL framework, when $\tau$ increased, the performance of Late Fusion also increased, whereas the performance of Conv Pooling decreased. This is intuitive, because more input information was ignored in Conv Pooling when $\tau$ increased.

The classification accuracies with DP post-processing are shown in the right panel of Table~\ref{tab:acc1}. Post-processing increased the classification accuracies for all classifiers and different $\tau$, e.g., the five classifiers achieved 2.3\%, 1.2\%, 2.1\%, 1.5\%, and 2.0\% performance improvement when $\tau=1$, respectively. However, as $\tau$ increased, the classification performance improvements became less obvious. After post-processing, the \emph{many-to-many} and \emph{one-to-many} frameworks had higher accuracies than the \emph{many-to-one} framework, and only \emph{many-to-many} consistently outperformed \emph{one-to-one} for all $\tau$, suggesting post-processing may be more beneficial when more input and output information was utilized.

\subsection{Root Mean Squared Error (RMSE)}

We also computed the root mean squared error (RMSE) between the true video label sequences and the classifications. The RMSEs without DP post-processing are shown in the left panel of Table~\ref{tab:rmse1}. All MTDL frameworks had lower RMSEs than the \emph{one-to-one} framework, suggesting again that using neighboring input or label information in multi-task learning was beneficial.

The results after DP post-processing are shown in the right panel of Table~\ref{tab:rmse1}. DP post-processing reduced the RMSE for all MTDL frameworks and different $\tau$, suggesting that DP was indeed beneficial. Though all MTDL frameworks outperformed the \emph{one-to-one} framework only at $\tau=1$, the \emph{many-to-many} framework consistently outperformed \emph{one-to-one} for all different $\tau$.

\begin{table*}[htbp] \center \linespread{1.35}\selectfont
\caption{RMSEs for different classification frameworks and $\tau$, before and after DP post-processing.}   \label{tab:rmse1}
\begin{tabular}{|c|c|c|c|c|c|c|c|} \hline
\multirow{2}{*}{Framework}   & \multirow{2}{*}{Method}          & \multicolumn{3}{c|}{RMSE without DP}                    & \multicolumn{3}{c|}{RMSE with DP}                     \\ \cline{3-8}
                             &                                  & $\tau=1$        & $\tau=4$        & $\tau=7$        & $\tau=1$        & $\tau=4$        & $\tau=7$        \\ \hline
One-to-one                   & ResNet50                         & 0.4840          & 0.4840          & 0.4840          & 0.4199          & 0.4199          & 0.4199          \\ \hline
\multirow{2}{*}{Many-to-one} & Conv Pooling                     & 0.4728          & 0.4690          & 0.4795          & 0.4066          & 0.4432          & 0.4419          \\ \cline{2-8}
                             & Late Fusion                      & 0.4761          & \textbf{0.4531} & 0.4740          & 0.4036          & 0.4254          & 0.4214          \\ \hline
One-to-many           & \multirow{2}{*}{Multi-Task Nets (ours)} & \textbf{0.4638} & 0.4695          & 0.4480          & \textbf{0.3964} & 0.4155          & 0.4260          \\ \cline{1-1} \cline{3-8}
Many-to-many                 &                                  & 0.4752          & 0.4640          & \textbf{0.4360} & 0.4085          & \textbf{0.4077} & \textbf{0.4083} \\ \hline
\end{tabular}
\end{table*}

\subsection{Training Time}

The training time of different models, averaged over five runs, is shown in Table~\ref{tab:time}. The training time of the \emph{many-to-one} and \emph{many-to-many} MTDL frameworks increased about linearly with the input context size; however, the training time of the \emph{one-to-many} MTDL framework was insensitive to $\tau$, which is an advantage.

\begin{table}[htbp] \center \linespread{1.35}\selectfont
\caption{Training time for different classification frameworks and $\tau$.} \label{tab:time}
\begin{tabular}{|c|c|c|c|c|} \hline
\multirow{2}{*}{Framework}      & \multirow{2}{*}{Method}  & \multicolumn{3}{c|}{Training time (s)}                  \\ \cline{3-5}
                                &                          & $\tau=1$              & $\tau=4$        & $\tau=7$      \\ \hline
One-to-one                      & ResNet50                 & \textbf{2231}         & \textbf{2231}   & \textbf{2231} \\ \hline
\multirow{2}{*}{Many-to-one}    & Conv Pooling             & 5318                  & 15378           & 29139         \\ \cline{2-5}
                                & Late Fusion              & 4892                  & 17390           & 27534         \\ \hline
One-to-many                     & \multirow{2}{*}{Multi-Task Nets (ours)} & 2246   & 2265            & 2542          \\ \cline{1-1} \cline{3-5}
Many-to-many                    &                          & 5759                  & 16182           & 27808         \\ \hline
\end{tabular}
\end{table}

\subsection{Comparison of Different Ensemble Approaches} \label{sect:ensemble_performance}

\begin{figure}[htpb]\centering
\subfigure[]{ \includegraphics[width=0.7\linewidth,clip]{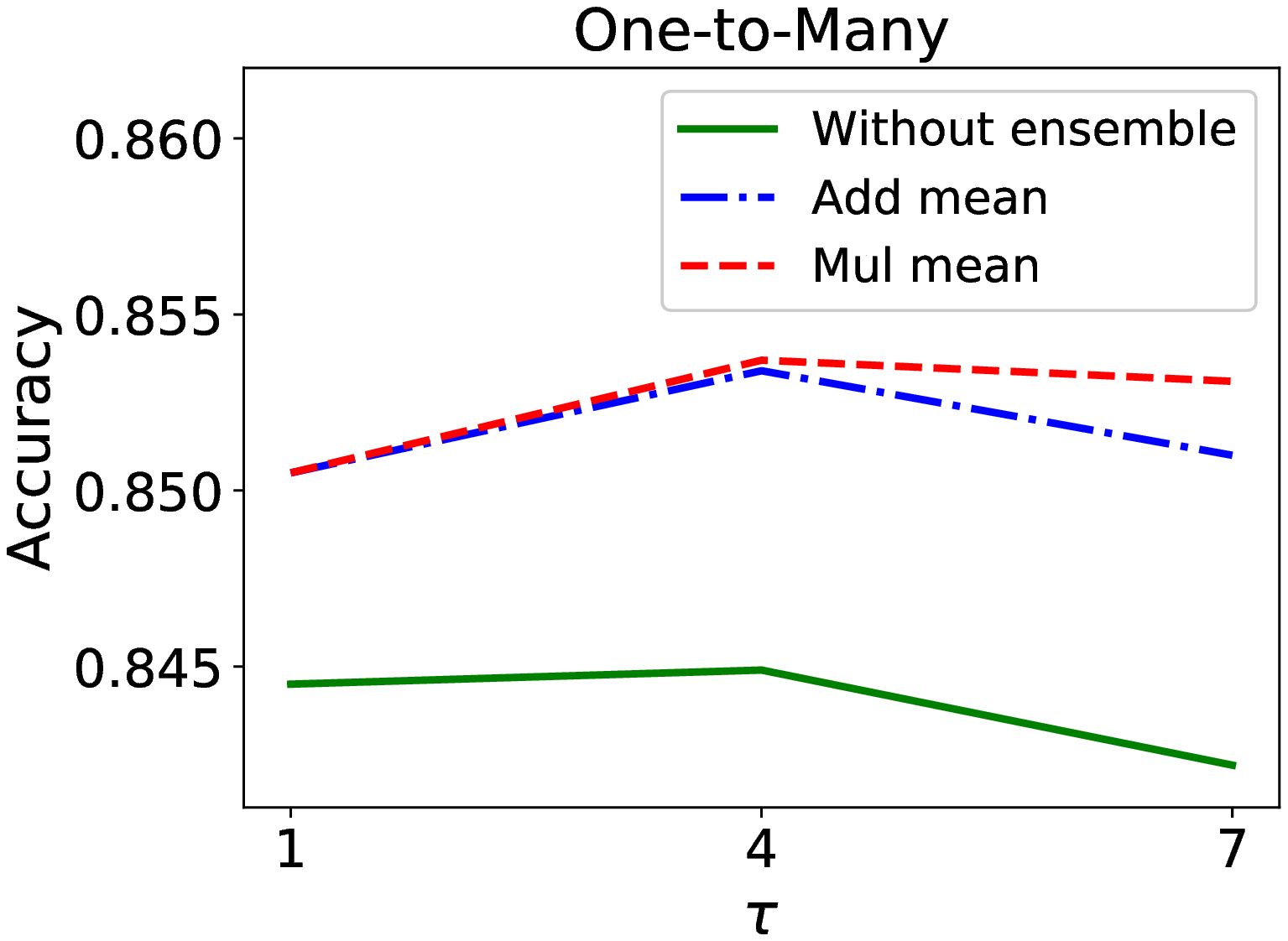}}
\subfigure[]{ \includegraphics[width=0.7\linewidth,clip]{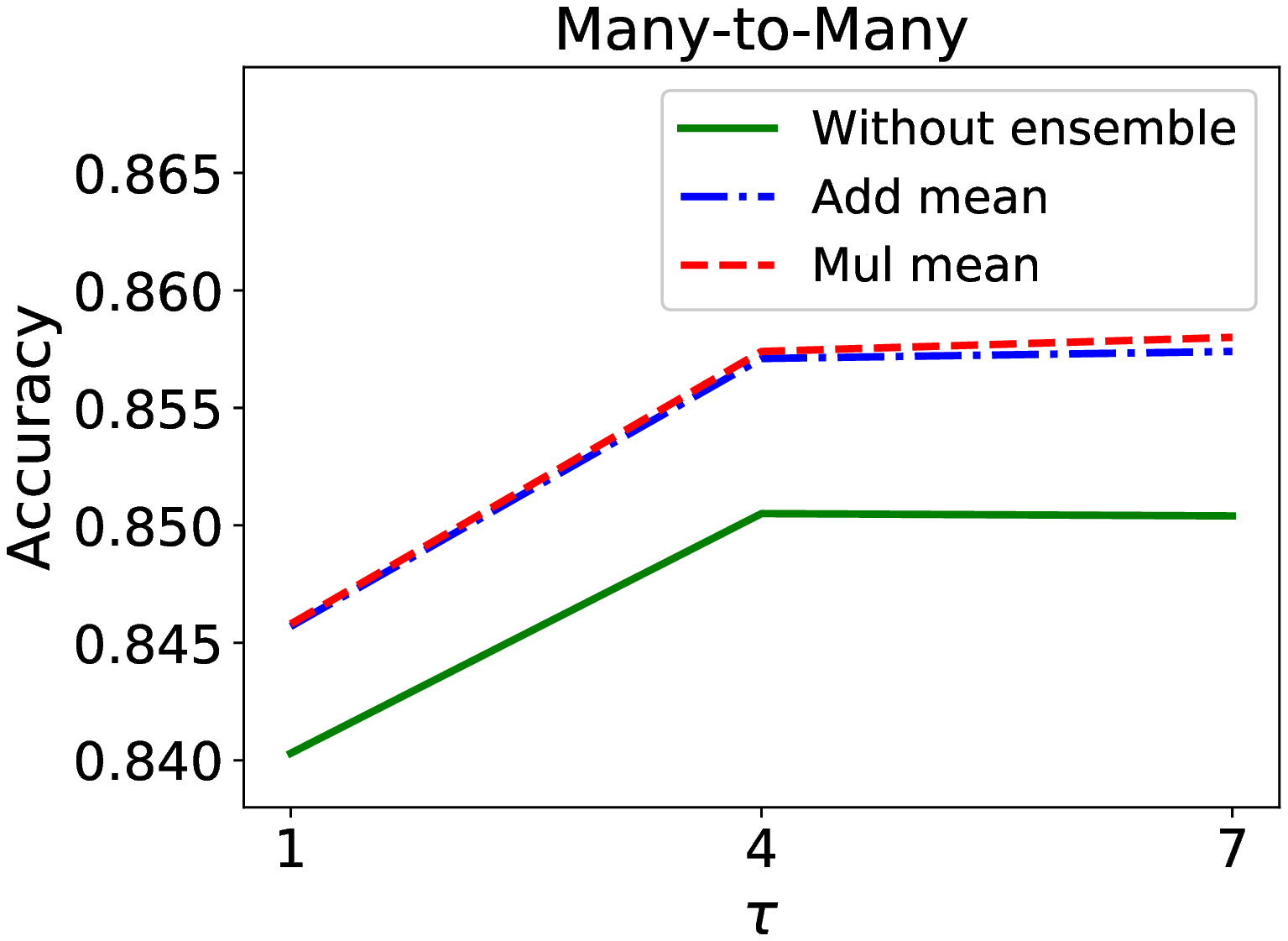}}
\caption{Classification accuracies with and without ensemble learning. (a) \emph{One-to-many}; (b) \emph{Many-to-many}. } \label{fig:ensemble_acc}
\end{figure}

\begin{figure}[htpb]\centering
\subfigure[]{ \includegraphics[width=0.7\linewidth,clip]{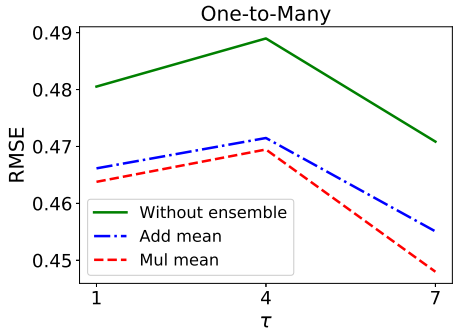}}
\subfigure[]{ \includegraphics[width=0.7\linewidth,clip]{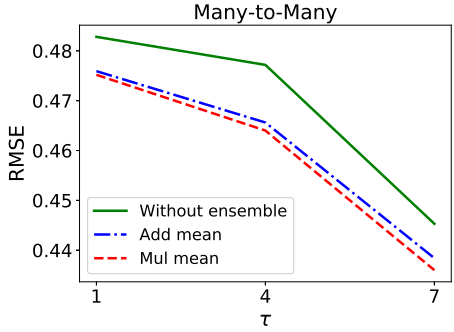}}
\caption{RMSEs with and without ensemble learning. (a) \emph{One-to-many}; (b) \emph{Many-to-many}. } \label{fig:ensemble_mse}
\end{figure}

We also compared the performances of different ensemble approaches introduced in Section~\ref{sect:ensemble}, without considering DP post-processing. The CNN models were constructed using the \emph{one-to-many} and \emph{many-to-many} MTDL frameworks. The results are shown in Figs.~\ref{fig:ensemble_acc} and \ref{fig:ensemble_mse}. Both \emph{additive mean} and \emph{multiplicative mean} achieved performance improvements. \emph{Multiplicative mean} also slightly outperformed \emph{additive mean}. As $\tau$ increased, the performance of the \emph{many-to-many} MTDL framework was improved. The \emph{one-to-many} MTDL framework had the best performance when $\tau=4$.

\subsection{Comparison of Different Losses in DP Post-Processing}

\begin{figure}[htpb]\centering
\subfigure[]{ \includegraphics[width=0.7\linewidth,clip]{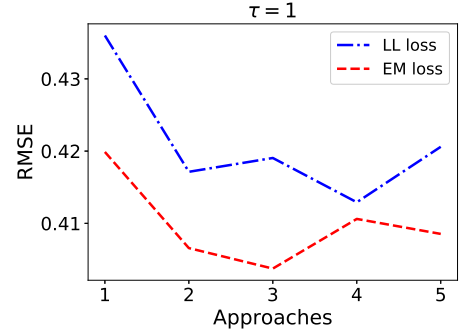}}
\subfigure[]{ \includegraphics[width=0.7\linewidth,clip]{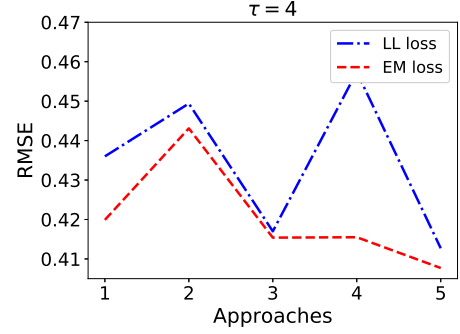}}
\subfigure[]{ \includegraphics[width=0.7\linewidth,clip]{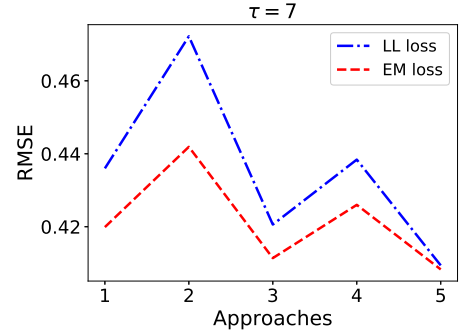}}
\caption{RMSEs of different per-frame losses in DP. (a) $\tau=1$; (b) $\tau=4$; (b) $\tau=7$. The numbers on the horizontal axis denote different MTDL frameworks: 1--\emph{One-to-one}, 2--\emph{Many-to-one} (Conv Pooling), 3--\emph{Many-to-one} (Late Fusion), 4--\emph{One-to-many}, 5--\emph{Many-to-many}. } \label{fig:dp_loss}
\end{figure}

Next, we studied the effect of different per-frame losses in DP post-processing. The RMSEs for different $\tau$ and different MTDL frameworks are shown in Fig.~\ref{fig:dp_loss}. The EM loss always gave smaller RMSEs than the LL loss.

\begin{figure}[htpb]\centering
\subfigure[]{ \includegraphics[width=0.7\linewidth,clip]{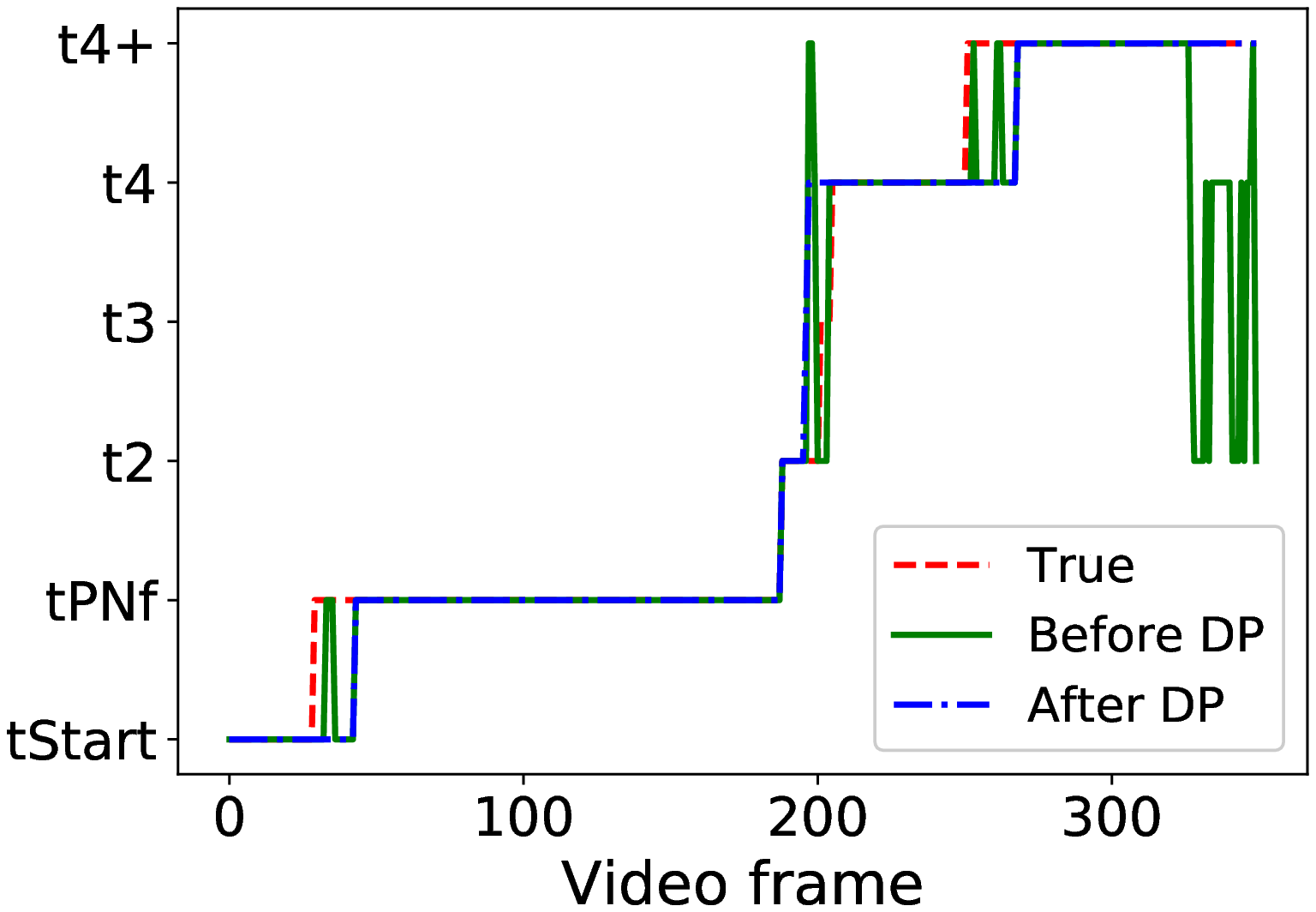}}
\subfigure[]{ \includegraphics[width=0.7\linewidth,clip]{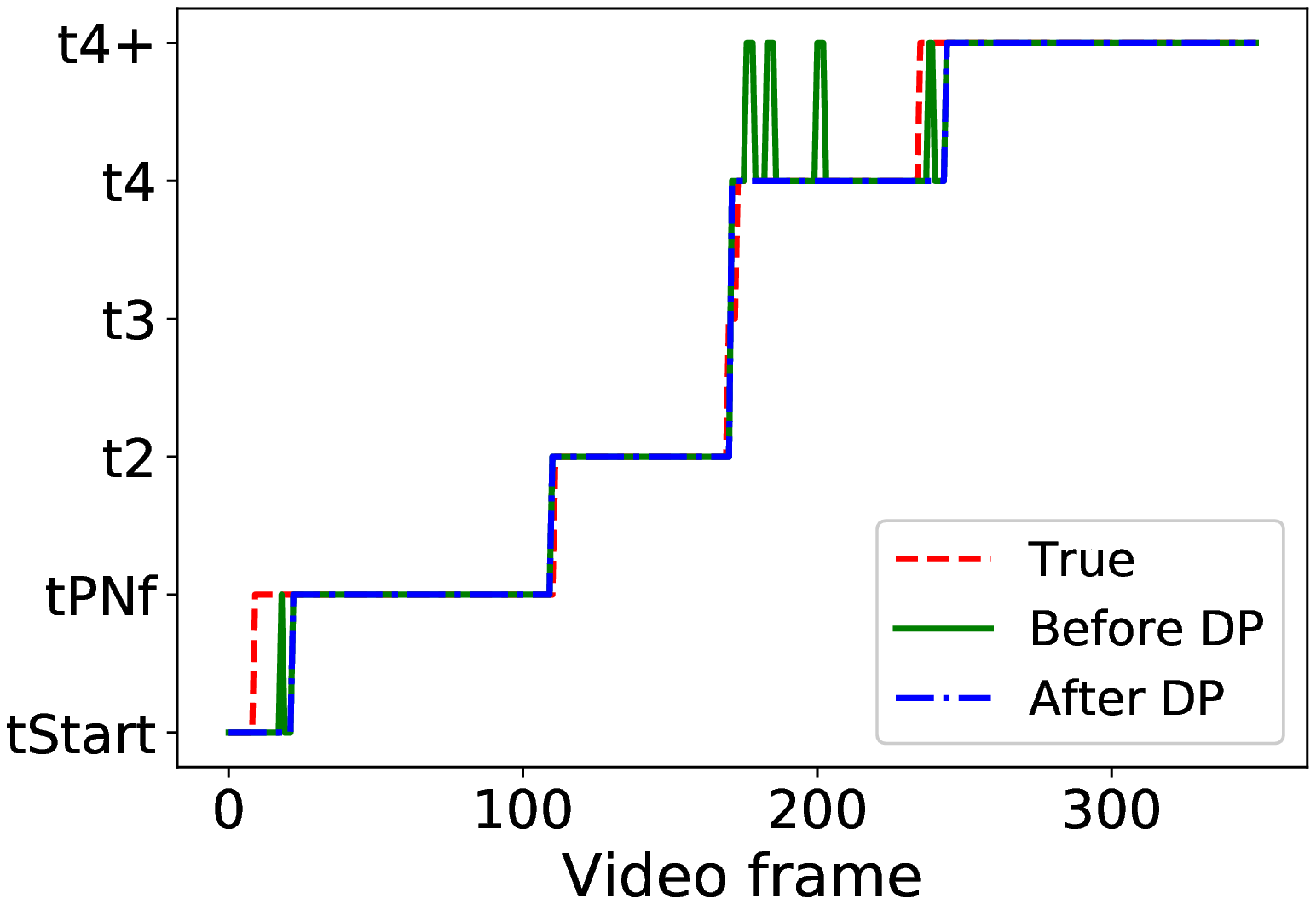}}
\caption{True stage labels, and classifications before and after DP, in two time-lapse videos. \emph{One-to-many} and $\tau=1$ were used.} \label{fig:cell_stage}
\end{figure}

The true stage labels, and the classified labels before and after DP in two time-lapse videos, are shown in Fig.~\ref{fig:cell_stage}. Clearly, DP smoothed the classifications, and its outputs were closer to the groundtruth labels.

\begin{figure}[htpb]\centering
\subfigure[\label{fig:cm_before}]{ \includegraphics[width=\linewidth,clip]{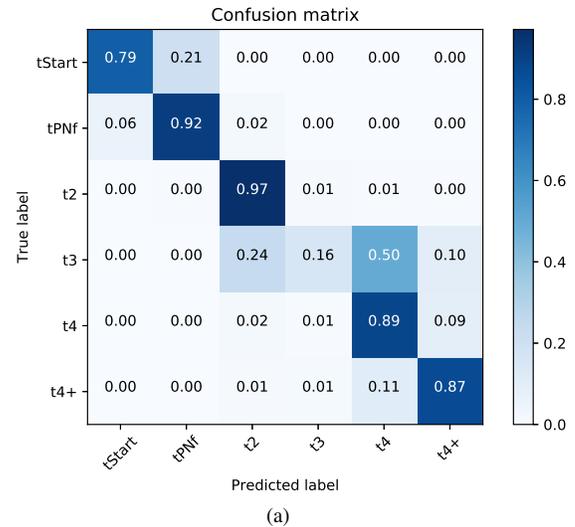}}
\subfigure[\label{fig:cm_after}]{ \includegraphics[width=\linewidth,clip]{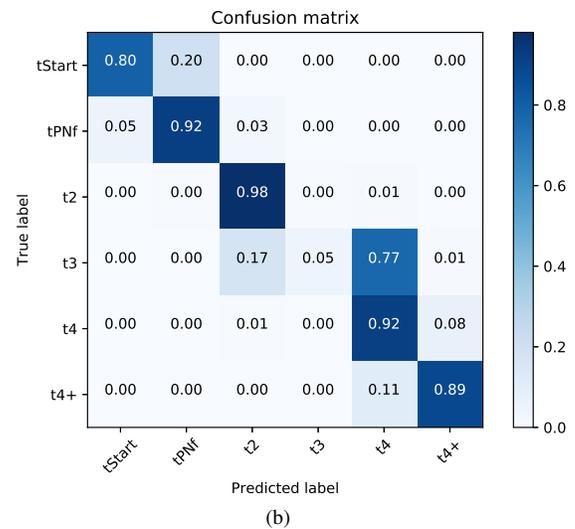}}
\caption{Confusion matrices (a) before and (b) after DP post-processing.} \label{fig:cm}
\end{figure}

The confusion matrix for the \emph{one-to-many} MTDL framework, using the \emph{multiplicative mean} and $\tau=1$, is shown in Fig.~\ref{fig:cm_before} before DP post-processing, and in Fig.~\ref{fig:cm_after} after DP post-processing. The diagonal shows the classification accuracy of each individual cell stage. Post-processing improved the accuracy of all embryonic stages except t3, whose classification accuracy before DP (16\%) was much lower than others. There may be two reasons for this: 1) Stage t3 had much fewer training examples in our dataset (see Fig.~\ref{fig:distribution}), and hence it was not trained adequately; and, 2) the low accuracy of t3 may also be due to multipolar cleavages from the zygote stage, which occurs in 12.2\% of human embryos  \cite{Kalatova2015}.

\section{Conclusion} \label{sect:conclusions}

Accurate classification of embryo early development stages can provide embryologists valuable information for assessing the embryo quality, and hence is critical to the success of IVF. This paper has proposed an MTDL-DP approach for automatic embryo development stage classification from time-lapse videos. Particularly, the \emph{one-to-many} and \emph{many-to-many} MTDL frameworks performed the best. Considering the trade-off between training time and classification accuracy, we recommend the \emph{one-to-many} MTDL framework in MTDL-DP, because it can achieve comparable performance with the \emph{many-to-many} MTDL framework, with much lower computational cost.

To our knowledge, this is the first study that applies MTDL to embryo early development stage classification from time-lapse videos.


\end{document}